\documentclass[12pt,preprint]{aastex}

\newcommand{\etal}{et al.}

\newcommand{\kev}{keV}
\newcommand{\fouru}{4U~1820-30}

\usepackage{emulateapj5}
\usepackage{times}
\usepackage{graphicx}

\begin{document}

\title{The evolution of the accretion disk around \fouru\ during a
  superburst}
\author{D.R. Ballantyne\altaffilmark{1} and
  T.E. Strohmayer\altaffilmark{2}}
\altaffiltext{1}{Canadian Institute for Theoretical
    Astrophysics, 60 St. George Street, Toronto, Ontario, Canada
    M5S~3H8; ballantyne@cita.utoronto.ca}
\altaffiltext{2}{NASA Goddard Space Flight Center, Laboratory
    for High Energy Astrophysics, Code 680, Greenbelt, MD, USA 20771;
  stroh@clarence.gsfc.nasa.gov}

\begin{abstract}
Accretion from a disk onto a collapsed, relativistic star -- a neutron
star or black hole -- is the mechanism widely believed to be
responsible for the emission from compact X-ray
binaries. Because of the extreme spatial resolution
required, it is not yet possible to directly observe the evolution or
dynamics of the inner parts of the accretion disk where general
relativistic effects are dominant. Here, we use the bright X-ray
emission from a superburst on the surface of the neutron star \fouru\
as a spotlight to illuminate the disk surface. The X-rays cause iron
atoms in the disk to fluoresce, allowing a determination of the
ionization state, covering factor and inner radius of the disk over
the course of the burst. The time-resolved spectral fitting shows that
the inner region of the disk is disrupted by the burst, possibly being
heated into a thicker, more tenuous flow, before recovering its
previous form in $\sim 1000$~s. This marks the first instance that the
evolution of the inner regions of an accretion disk has been observed
in real-time.
\end{abstract}


\section{Introduction}
\label{sect:intro}
Superbursts \citep[e.g.][]{cor00,sb01,kuul02} are the more powerful and rarer
relations of the Type~I X-ray bursts, and are thought to be due to the
nuclear burning of the ashes (principally carbon and rp-process
nuclei) that remain after thermonuclear processing of the accreted
light elements \citep{sb01,cumm03}. One of the defining characteristics
of a superburst is its long decay time of a number of hours,
three orders of magnitude greater than a typical Type~I
burst \citep{kuul03}. This allows high signal-to-noise spectra to be
accumulated at regular intervals throughout the course of the burst.

\fouru\ is a compact low-mass X-ray binary with a period of only
11.4~minutes \citep{spw87}, and lies within the globular cluster
NGC~6624 at a distance of $\sim 7$~kpc
\citep{vlv86,hs85,rml93}. Type~I X-ray bursts have been observed from
this system for nearly 30 years \citep{gg76}. The accretion rate
inferred from the persistent X-ray luminosity is $\sim 1.5\times
10^{17}$~g~s$^{-1}$ \citep{cumm03}. The first, and so far only,
superburst seen from \fouru\ was observed by the \textit{Rossi X-ray Timing
  Explorer} (\textit{RXTE}) on 1999 September 9 (UT) during a scheduled
monitoring observation. The initial analysis and interpretation of the
data was presented by \citet{sb01}. Time-resolved spectroscopy
presented by these authors showed that an iron K$\alpha$ line and absorption
edge were present in the spectra, and that their properties changed
over time. A likely origin for these features is reprocessing of the
X-rays by the accretion disk \citep{dd91}. This process, called X-ray
reflection, is commonly used to explain similar features in the
spectra of Seyfert galaxies \citep*[e.g.][]{pou90,gf91,bal03}.

In this paper we present the results of fitting the time-resolved
spectra of the superburst with detailed models of X-ray
reflection. The next section describes the model calculations and
details of the spectral fitting. The results are presented in
Section~\ref{sect:res} and discussed in Section~\ref{sect:discuss}.

\section{Model Calculations}
\label{sect:models}
The X-ray reflection models were calculated with the code of
\citet{ros93} (see also \citealt*{ros99,bif01,brf01}). A
one-dimensional slab of uniform density gas is illuminated by a
blackbody continuum (defined from 1~eV to 100~\kev) of temperature
$kT$, and allowed to come into thermal and ionization balance. The
emergent spectrum is calculated with a Fokker-Planck/diffusion method
that naturally accounts for the energy redistribution of photons by
Compton scattering \citep*{rwm78}. The following ions are included in the
calculations: \ion{C}{5} \ion{--}{7}, \ion{N}{6} \ion{--}{8},
\ion{O}{5} \ion{--}{9}, \ion{Mg}{9} \ion{--}{13}, \ion{Si}{11}
\ion{--}{15} and \ion{Fe}{16} \ion{--}{27}. Fluorescence and
recombination lines from these ions are computed and transferred
simultaneously with the continuum, resulting in a self-consistent
prediction of the emitted spectrum.

The compact orbit of this binary suggests that the donor is an evolved
low-mass helium star \citep{rap87}, and thus the accretion disk
material will be He-rich and H-poor \citep{cumm03}. Therefore, we
assumed the reflecting medium has the He, C, N, and O abundances of
cool extreme helium stars from the tabulation of \citet{pan01}. The
Mg, Si and Fe abundances were left at their solar values
\citep{gns96}, but the effective metallicity is increased (by $\sim
3$) due to the negligible levels of hydrogen. The helium is assumed to
be completely ionized everywhere in the reflecting region.

The computed reflection spectra are parameterized by the ionization
parameter $\xi=4 \pi F_{\mathrm{X}} / n_{\mathrm{He}}$ of the
irradiated gas, where $F_{\mathrm{X}}$ is the flux of the incident
X-rays and $n_{\mathrm{He}}=10^{15}$~cm$^{-3}$ is the assumed number
density of helium. Since the blackbody emission is also seen directly
from the neutron star, the fitted spectrum is the sum of the blackbody
and the reflected emission, weighted by a factor $R$, the reflection
fraction. This parameter measures the relative importance of
reflection in the spectrum, and can be related to the covering factor
of the accretion disk.

\citet{sb01} describes the \textit{RXTE} data and the process of
accumulating the $\sim 80$ 64~s spectra during the burst. The preburst
spectrum was used as the background, and was thus subtracted from the
data during spectral fitting \citep{sb01}. This persistent emission
(amounting to a luminosity of $1.8\times 10^{37}$~erg~s$^{-1}$) is
driven by the accretion onto the neutron star, so using this spectrum
as a background implicitly assumes that the burst will not effect this
underlying emission. As we see below, we cannot be sure that this does
not happen; however, \citet{sb01} found little difference in their
spectral fits between the preburst background and the detector
background. Since the flux of the superburst varies from 20$\times$ to
3$\times$ greater than the preburst emission, then the shape of the
background will not greatly effect the fit results.

The spectral fitting was performed with \textsc{xspec}v.11.2.0bp
\citep{arn96}. The model consisted of a grid of reflection spectra
(read in using the \texttt{atable} command) and attenuation from
neutral gas along the line of sight (using the \texttt{wabs} model),
parameterized by the column density $N_{\mathrm{H}}$. The Galactic
absorption toward \fouru\ is $1.5\times 10^{21}$~cm$^{-2}$
\citep{dl90}. The reflection spectra were blurred by the effects of
relativistic motion within the Schwarzschild metric \citep{fab89}. The
inner radius of emission ($r_{\mathrm{in}}$) was allowed to vary, but
the emissivity was fixed as $r^{-3}$, where $r$ is distance along the
accretion disk, and the outer radius was frozen at $200$~$GM/c^{2}$,
where $M$ is the mass of the neutron star. The inclination angle of
the disk to the line of sight was fixed at 30$^{\circ}$. The fitted
energy range was 3--40~\kev, but was changed to 3--15~\kev\ 7000~s
into the burst because of background dominating at higher
energies. The errorbars on the best fit parameters were computed using
the 2$\sigma$ uncertainties for one parameter of interest.

\section{Spectral Fitting Results}
\label{sect:res}
The results of fitting the reflection models to the superburst data
are shown in Figure~\ref{fig:fitres}. We find that an ionized
reflector provides a good description to the spectra for most of the
burst. During the first 1000~s the spectra are well fit with a large
$\xi$, indicating significant ionization of the accretion disk. Over
the next $\sim 500$~s $\xi$ decreases by more than an order of
magnitude. For times between $\sim 2500$ and $3000$~s the spectra
cannot be fit by the reflection model, except for the last three
points, where only very loose constraints could be placed on
$\xi$. \citet{sb01} showed the spectrum was hardening very rapidly
during that period, which indicates that some rapid evolution of the
system was ongoing, and is not well modeled by our equilibrium
calculations. However, the recombination timescale for H-like Fe is
$t_{\mathrm{recomb}} \sim 1/n_{e}\alpha_{\mathrm{Fe}}(T)$. The slowest
timescale occurs at the highest temperature, which is $10^7$~K at the
surface of the reflector, so that $t_{\mathrm{recomb}} \sim
10^{-4}$~s, assuming the recombination coefficient of
\citet{vf96}. This is the same order of magnitude as the disk
dynamical time at 10~$GM/c^2$. The recombination timescale of
\ion{He}{2} is $\sim 0.1$~s, much longer than the dynamical time, but
faster than the 64~s spectra examined here. Therefore, the assumptions
of ionization equilibrium or fully stripped helium cannot be the cause of the poor fits in this
region of the lightcurve. A more fundamental change in the X-ray
source or reflecting geometry may be required. Finally, after the last
Earth occultation passage, the system was revealed to have settled
into a final decaying state, with $\xi \sim 100$~erg~cm~s$^{-1}$,
although one other non-equilibrium event took place at $\sim 6000$~s.

While the disk inclination angle was fixed at 30$^{\circ}$ for the
spectral fits, there is a previous constraint of
35$^{\circ}$--50$^{\circ}$ found by \citet{and97} from modulations in
the ultraviolet flux (due to X-ray heating of the secondary). Relativistic disk
features are also dependent on the accretion disk inclination angle
\citep{fab89}, so can provide a consistency check on the reflection
hypothesis.  In Figure~\ref{fig:incl} we plot the 68\%, 90\% and 99\%
confidence contours for $r_{\mathrm{in}}$ and the disk inclination
angle 5500~s into the superburst. This segment was chosen as it has a
low reduced $\chi^2$ and strong reflection features. While the disk
fit seems to prefer a lower inclination angle, it is consistent with
the results of \citet{and97} at the 2$\sigma$ level.

The observed evolution of $\xi$ from highly to weakly ionized is to be
expected since the observed flux decays over the course of the burst
\citep{sb01}. More fundamental insights from this analysis arise from
the variations of $N_{\mathrm{H}}$, $R$ and $r_{\mathrm{in}}$ over the
first 2000~s of the superburst.  The changes of both
$r_{\mathrm{in}}$, which measures the inner radius of the disk (as
judged from the width of the iron line and edge), and $R$, the
reflection fraction, strongly indicates that the accretion disk is
significantly affected by the superburst out to a radius of $\sim
100$~$GM/c^2$. In addition, the line-of-sight absorbing column also
went through a similar variation, either as a result of the burst
itself or the changes to the inner accretion disk. All three of these
parameters follow a similar pattern where they increase
($r_{\mathrm{in}}$ and $N_{\mathrm{H}}$) or decrease ($R$) before
returning to close to their initial value. This behavior seems to be
closely linked to the blackbody temperature $kT$, rather than the
flux, which decreases monotonically during the burst.

\section{Discussion}
\label{sect:discuss}
A likely scenario to explain the observed behavior is that the
accretion flow in the inner disk is altered by the superburst so that
it loses the ability to produce reflection features. This could be
accomplished if the scale-height of the disk, $H$, increased due to
heating of the gas by the radiation field. This would decrease the
surface density of the disk, allowing it to be more easily and
completely ionized by the burst radiation, removing any spectral
features in the observable band, and effectively becoming a mirror for
the blackbody spectrum. As there would be no evidence for reprocessing
from this innermost region, the estimate of the inner radius would be
from further out in the disk where iron can still produce an emission
line. The additional blackbody emission reflected from the ionized
mirror, would dilute the reflection features, resulting in the low
value of $R$. Alternatively, the lower surface density could reduce
the optical depth of the gas to the point where the inner disk becomes
a very inefficient reflector, while only the outer disk would be
optically thick enough for efficient reprocessing. 

The main evidence for this thermal effect is seen in the correlations
between $r_{\mathrm{in}}$, $R$ and $kT$. The Compton temperature of
the gas is $kT_{\mathrm{C}} \approx kT$, and reached $\sim 3$~\kev,
about twice the maximum disk temperature \citep{ss73}. Since $H
\propto c_{\mathrm{s}}r^{3/2} \propto T^{1/2} r^{3/2}$, where
$c_{\mathrm{s}}$ is the local sound speed, then $H$ will increase by
$\sim 40$\% from its non-illuminated value. Changes to the disk
surface density can take place over a viscous time which is $\sim
1000$~s at 100~$GM/c^{2}$ if $\alpha \sim 0.2$, where $\alpha$ is the
Shakura-Sunyaev viscosity parameter \citep{ss73}. This timescale
compares favorably with the observed changes in $R$ and
$r_{\mathrm{in}}$. An increase in $H$ may also explain the evolution
of $N_{\mathrm{H}}$ as the thicker disk may bring clumps of disk
material into the line-of-sight. Data which extend below the 3~\kev\
limit of \textit{RXTE} would be needed to test if the absorption was
caused by ionized or neutral material. Under this hot disk scenario,
once the peak in $kT$ is passed the disk material cooled to lower and
lower temperatures, and thus contracted to its original
size. Simultaneously, the amount of material available to reflect
increased, so $R$ grew larger, but $\xi$ dropped due to a combination
of the increasing surface density and the decaying burst flux.

The connection between the ionization parameter and the disk surface
density $\Sigma$ can be formulated as $\xi \approx L_{\mathrm{X}}
m_{\mathrm{He}} H / r^2 \Sigma$, where $L_{\mathrm{X}}$ is the
luminosity of the burst, and $m_{\mathrm{He}}$ is the mass of a helium
nucleus. Using the equations for a gas pressure dominated
Shakura-Sunyaev disk with electron scattering opacity \citep{ss73}, a
peak burst luminosity of $3.4 \times 10^{38}$~erg~s$^{-1}$ \citep{sb01}, and
$\alpha=0.2$, we find $\xi \approx 100$~erg~cm~s$^{-1}$ for
$r=15$~$GM/c^2$, which is consistent with the values of $\xi$ found
near the end of the burst (Fig.~1). From the observed decay in
flux \citep{sb01}, the surface density of the disk must be $\sim 100$
times less than the Shakura-Sunyaev value at the start of the
superburst, but only $\sim 6$ times less near the end in order to
obtain the necessary level of ionization. This indicates the surface
density of the inner accretion disk was significantly lower at the
peak of the burst but increased as the burst decayed, supporting the
interpretation that the disk material was severely affected by the explosion.

Another process that could lower the surface density at the inner edge
of an accretion disk is the interaction between the radiation field
and material flowing in the disk. Radiation can remove angular
momentum from the disk material, causing the disk to empty inward,
thus lowering the surface density. There exists only a few
calculations of the interaction between an accretion disk and the
radiation field from a luminous central source
\citep{wm89,walk92,ml93,ml96}. These models predict that the radiation
will exert a significant torque on the accretion flow, which increases
the accretion rate close to the central object. This effect may be
enhanced in the case of \fouru\ due to the power of the superburst (as
compared to the typical Type~I burst considered in these models), as
well as an interaction between the disk and the expanding shell that
was ejected from the neutron star just prior to the superburst
\citep{sb01} -- the radius-expansion phase \citep{lvb84}. It is
difficult to rigorously compare this idea to the observations because
detailed modeling of the disk-radiation interaction for a superburst
or a radius-expansion burst has not been performed.

By the end of the observation, the accretion disk has recovered from
its disruption by the burst, and is reflective down to a radius of
10--20~$GM/c^2$, which is curiously larger than the innermost stable
circular orbit of 6~$GM/c^2$. Since this is not an X-ray pulsar, the
disk would not be truncated by the star's magnetosphere. Rather, this
may be evidence for the disk-radiation interaction described above,
and the material within 10--20~$GM/c^2$ is no longer dense enough to
provide a reflection signature. At these late times, the fits also
require strong reflection, with $R>1$, indicating that the disk covers
a large solid angle from the X-ray source, and thus is thin relative
to the radius of the neutron star or has a non-negligible flare at
large radii. The latter possibility is supported by the reflection
from $r > 100$~$GM/c^2$ observed early in the burst. If the inner disk
expanded from the X-ray heating, then a significant disk flare might
be necessary in order to allow the distant material to be irradiated
by the blackbody emission. 

\begin{figure}
\centerline{
\includegraphics[angle=-90,width=0.7\textwidth]{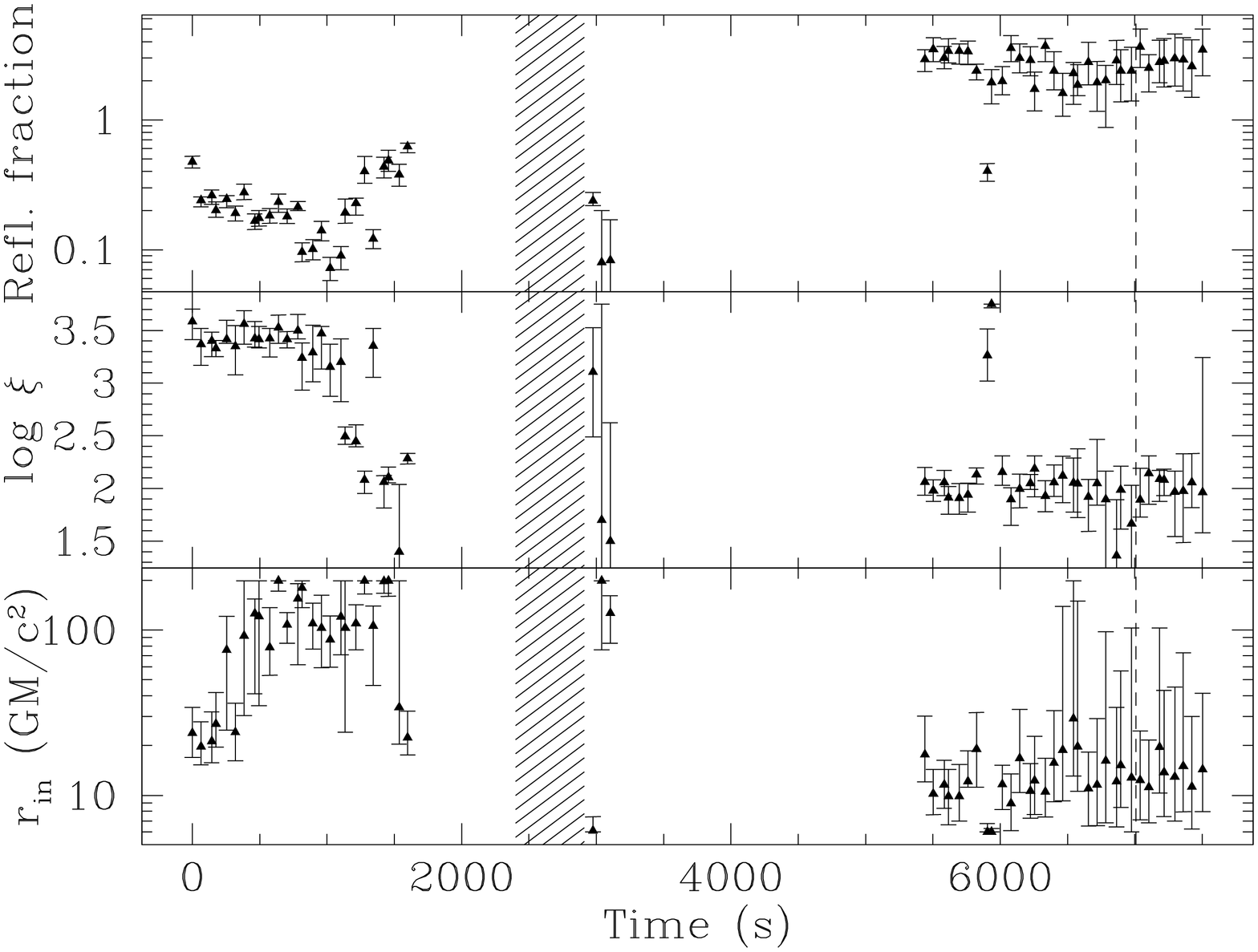}
}
\centerline{
\includegraphics[angle=-90,width=0.7\textwidth]{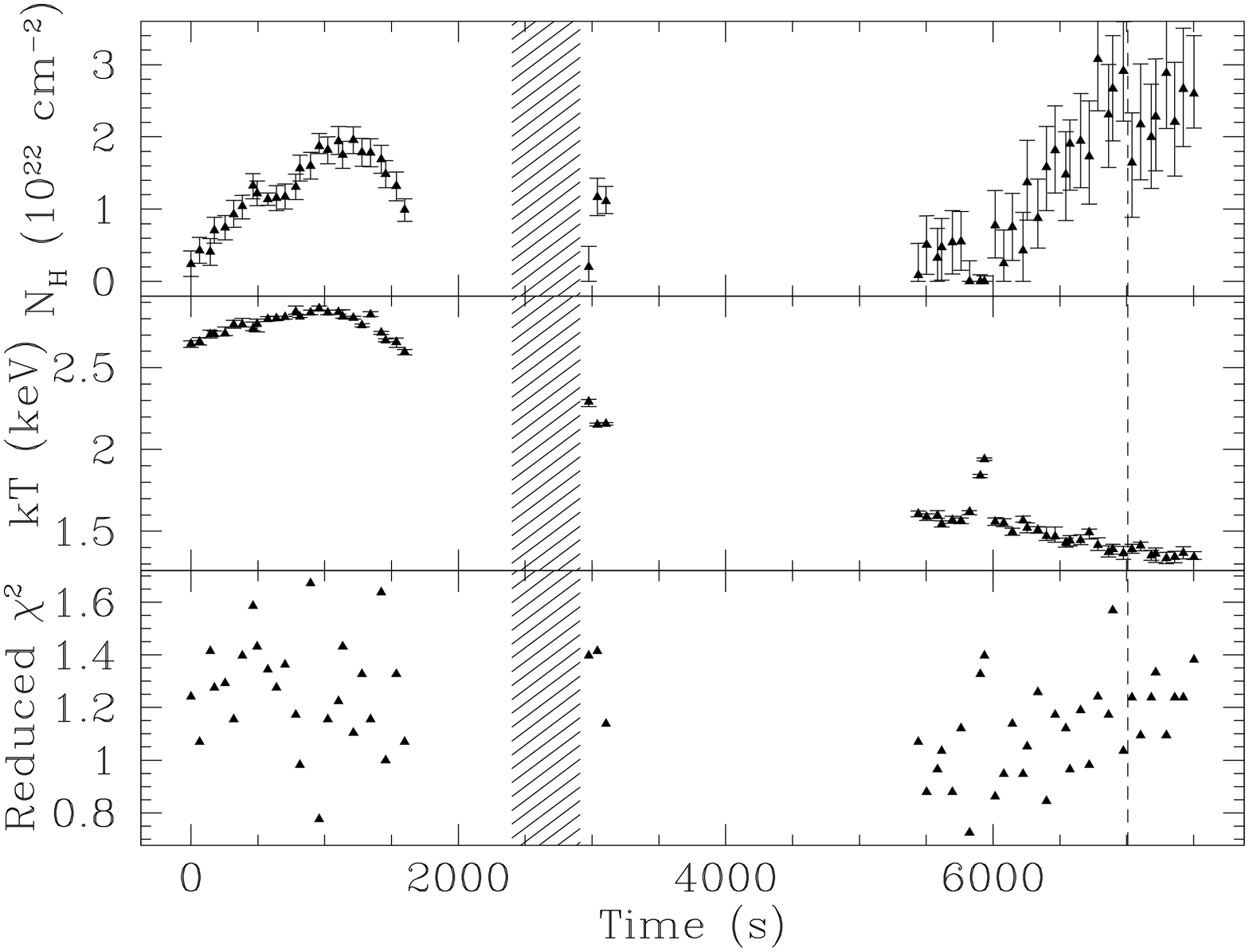}
}
\caption{\footnotesize Results of fitting the time-resolved superburst spectra of
\fouru\ with ionized reflection models. The top panel shows the
evolution of the reflection parameters: the ionization parameter
$\xi$, the reflection fraction, and inner radius of the disk,
$r_{\mathrm{in}}$. The lower panel describes the evolution of the
absorption column, $kT$ of the incident blackbody, and the reduced
$\chi^2$ of the spectral fits. The reflection fraction,
$r_{\mathrm{in}}$, and $N_{\mathrm{H}}$ seem to be closely related to
the evolution of $kT$. Data were fit from 3--40~\kev\ for the points
to left of the dashed line, and from 3--15~\kev\ for the points to the
right. Time equals zero corresponds to about 100~s after the start of
the burst. The hatched region denotes the interval where the spectra
were hardening rapidly, and were not well described by the equilibrium
reflection model.}
\label{fig:fitres}
\end{figure}

\begin{figure}
\centerline{
\includegraphics[angle=-90,width=0.8\textwidth]{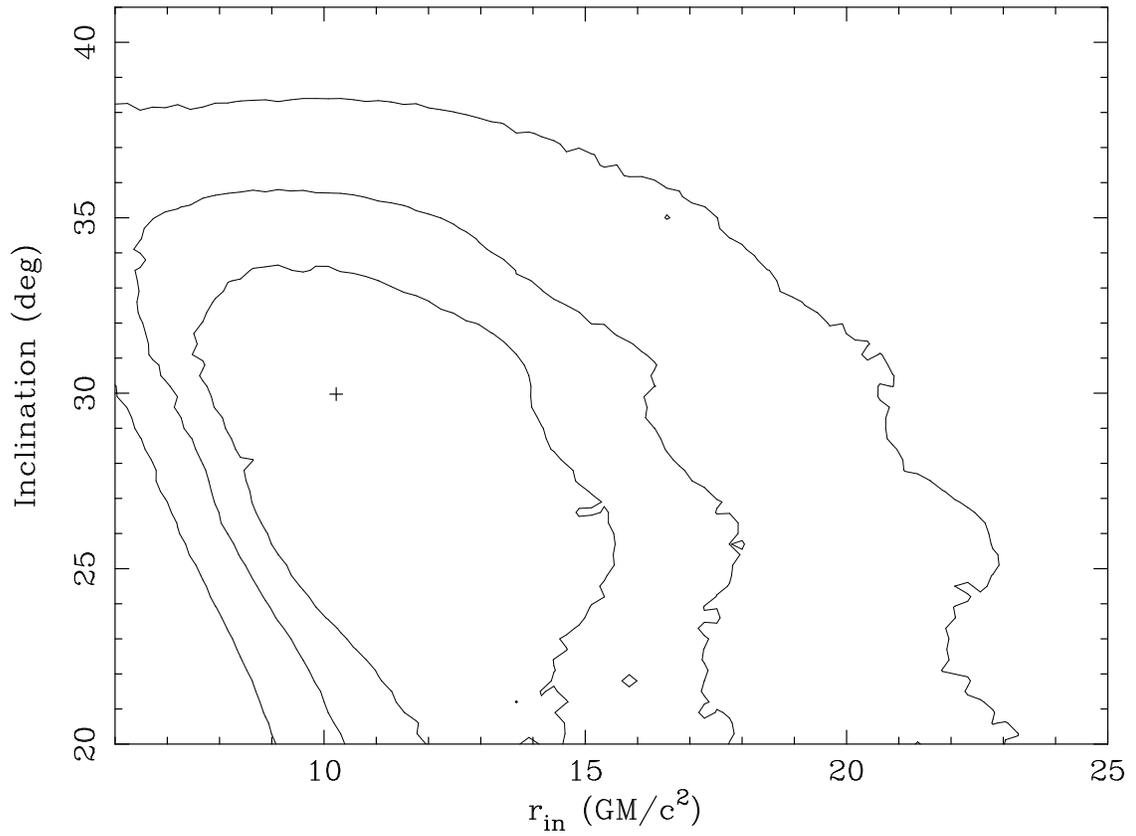}
}
\caption{Joint 68\%, 90\% and 99\% confidence contours on
$r_{\mathrm{in}}$ and the disk inclination angle for a spectral fit
5500~s into the superburst. Modulations in the ultraviolet flux from
\fouru\ previously constrained the inclination angle to between
$35^{\circ}$ and $50^{\circ}$ \citep{and97}, which is consistent with
the results from disk reflection at the 2$\sigma$ level.}
\label{fig:incl}
\end{figure}

Dipping behavior is seen in the X-ray lightcurve of \fouru\ between
6000--7500~s into the burst (see Fig.~7 in \citealt{sb01}). This kind
of variability in low-mass X-ray binaries is often explained as
variable absorption from a two-component ionized medium close to the
neutron star \citep{fkl87}. The dipping observed from \fouru\ could
then be due to the reformation of the accretion disk, however, the
timescale is incorrect, with the dipping seen $\sim 2000$~s after the
probable recollapse of the disk. However, there is an interesting
correspondence between the times when dipping is observed and an
increase in $N_{\mathrm{H}}$ derived from spectral fitting
(Fig.~\ref{fig:fitres}). This may indicate that the dipping is caused
by partially ionized material that is slowly recombining, but farther away
from the accretion disk. An obvious source for such material is the
gas that was cast off from the neutron star during the radius
expansion phase. This material could be at a large distance from the
system ($10^{14}$~cm if ejected at the escape velocity).

\acknowledgments

We thank E. Agol, L. Bildsten, E. Quataert, N. Murray \& C. Matzner
for helpful discussions, and A. Fabian for providing the relativistic
blurring functions. DRB is supported by the Natural Sciences and
Engineering Research Council of Canada.

\clearpage

\clearpage

\end{document}